\documentclass[twocolumn,prb,showpacs]{revtex4}
\usepackage{graphicx}
\usepackage{dcolumn}
\usepackage{amsmath}

\begin{document}

\newcommand{\K}{{\vec k}}

\title{Electronic phase transitions in the half-filled ionic Hubbard model}

\author{L. Craco,$^1$ P. Lombardo,$^2$ R. Hayn,$^2$ G.I. Japaridze,$^3$}
\author{E. M\"uller-Hartmann$^4$}
\affiliation{$^1$Max-Planck-Institut f\"ur Chemische Physik fester Stoffe,
01187 Dresden, Germany \\
$^2$Laboratoire Mat\'eriaux et Micro\'electronique de Provence associ\'e au
Centre National de la Recherche Scientifique. UMR 6137.
Universit\'e de Provence, France \\
$^3$ Andronikashvili Institute of Physics, Tamarashvili 6, 0177
Tbilisi, Georgia \\
$^4$Institut f\"ur Theoretische Physik,
Universit\"at zu K\"oln, Z\"ulpicher Strasse 77, 50937 K\"oln, Germany}

\date{\today}

\begin{abstract}
A detailed study of electronic phase transitions in the ionic Hubbard
model at half filling is presented. Within the dynamical mean field
approximation a series of transitions from the band insulator via a
metallic state to a Mott-Hubbard insulating phase is found at
intermediate values of the one-body potential $\Delta$ with increasing
the Coulomb interaction $U$. We obtain a critical region in which the
metallic phase disappears and a {\it novel} coexistence phase between
the band and the Mott insulating state sets in. Our results are
consistent with those obtained at low dimensions, thus they provide
a concrete description for the charge degrees of freedom of the ionic
Hubbard model.
\end{abstract}


\maketitle


One important problem in the field of electronic structure theory is
the understanding of the Mott metal-insulator transition
(MIT).~\cite{Mott,Imada} Many details have already been clarified
using the canonical model for this transition --- the one-band
Hubbard model. At half filling (i.e., having in average one electron
at each lattice site) this transition from the metallic to the
Mott-Hubbard insulating (MI) phase occurs with increasing the
on-site repulsion $U$. Great progress in understanding the MIT in
Hubbard-like models has been achieved in the last decade with the
development of the dynamical mean field theory (DMFT). Although many
aspects of the coherent (Fermi liquid) metallic phase and the
incoherent MI phase are now quite well understood, other questions
remain open, especially the relationship between the MI regime and
band insulators (BI), where one sub-band is almost completely filled
by two electrons such that an excitation gap is formed.

To study the interplay between band and Mott-Hubbard insulators,
various extended versions of the one-band Hubbard model have been
proposed. It is remarkable that different models show separate kinds
of behavior. Some of them have a crossover from the BI to the MI
regime,~\cite{FuRo} whereas others have clear transition points with
a metallic phase in between.~\cite{lc} A widely studied model for
the second class is the ionic Hubbard model
(IHM).~\cite{Garg,Paris,Kancharla} In the one-dimensional (1D) IHM,
however, a ferroelectric (or bond-ordered) phase is realized which
separates BI and MI, and the metallic phase shrinks to only one
point.~\cite{Fabrizio,Kampf} Already in 1D, a finite metallic region
can be recovered by introducing an intra-sublattice hopping
$t^{\prime}$ into the IHM.~\cite{Japaridze} In 2D~\cite{Kancharla}
or at larger dimensions~\cite{Garg} a correlation induced metallic
phase was reported between BI and MI, but it is still under debate
whether this metallic phase shrinks to a line~\cite{Garg} or if it
ends up at certain particular point by increasing the ionicity
$\Delta$. In this work we show that the metallic phase disappears at
a certain value of $\Delta$ above which we found a coexistence
region composed of distinct insulating phases.

It might be not very surprising to find quite different behavior in
various models at different dimensions. But even if we restrict our
attention to the IHM, the phase diagram is highly disputed. The 2D
case was studied by a cluster extension of DMFT~\cite{Kancharla} and
by the determinant quantum Monte Carlo method~\cite{Paris,Bouadim} showing
different results. The cluster-DMFT study suggests a bond-order
phase separating BI and MI regimes, while the stability of the
metallic phase was established on a 2D square lattice at finite
temperatures.~\cite{Paris}

To clarify the many open questions related to the electronic phase
diagram of the IHM, we study this problem here in the limit of high
dimensions. In this limit the self-energy is ${\bf k}$-independent
and, thus, the dynamical many-body problem can be mapped onto a
single-site one. Using basically the same DMFT scheme as Garg {\it
et al.},~\cite{Garg} we confirm the existence of the metallic phase
for moderate values of $U/\Delta$, but we find also important new
features which are different from those reported in this recent
work. So, for intermediate values of $\Delta$, we find a continuous
BI-to-metal phase transition and a discontinuous one between the
metallic and the MI phase with increasing the Coulomb interaction
$U$.

We start our study with the IHM
\begin{eqnarray}
H=& &-t\sum_{i \in A, j \in B , \sigma} ({\hat c}^+_{i\sigma} {\hat
c}_{j\sigma}+
H.c.) +U\sum_{i}{\hat n}_{i\uparrow}{\hat n}_{i\downarrow} \nonumber \\
& + & \Delta \sum_{i \in A} {\hat n}_{i} - \Delta \sum_{i \in B}
{\hat n}_{i}
\label{hamiltonien}
\end{eqnarray}
on a bipartite lattice with two sublattices $A$ and $B$ having
different on-site energies $\varepsilon_A=+\Delta$ and
$\varepsilon_B=-\Delta$. $t$ is the (inter-sublattice) hopping term
and $U$ is the on-site Coulomb repulsion. ${\hat c}^+_{i\sigma}$
(respectively ${\hat c}_{i\sigma}$) denotes the creation
(respectively annihilation) operator of an electron at the lattice
site $i$ with spin $\sigma$ and ${\hat n}_{i\sigma}$ is the
occupation number operator for electrons with spin $\sigma$. We
consider an average number of one particle per lattice site, i.e.,
the half filled case.

Application of the DMFT scheme to the IHM implies one to solve two $(A,~B)$
local problems given by the local one-particle Green's functions $G_\alpha$.
At high dimensions, these local propagators are connected self-consistently
by the Dyson equation
${\mathcal G}_{0\alpha}^{-1}=G_\alpha^{-1} + \Sigma_\alpha$.~\cite{Georges}
Without loss of generality, in what follows we will perform our calculations
using a semi-circle density of states (DOS) --- this type of one-particle DOS
is realized on a Bethe lattice with large coordination number.

\begin{figure}[t]
\begin{center}\leavevmode
\includegraphics[width=8.0cm]{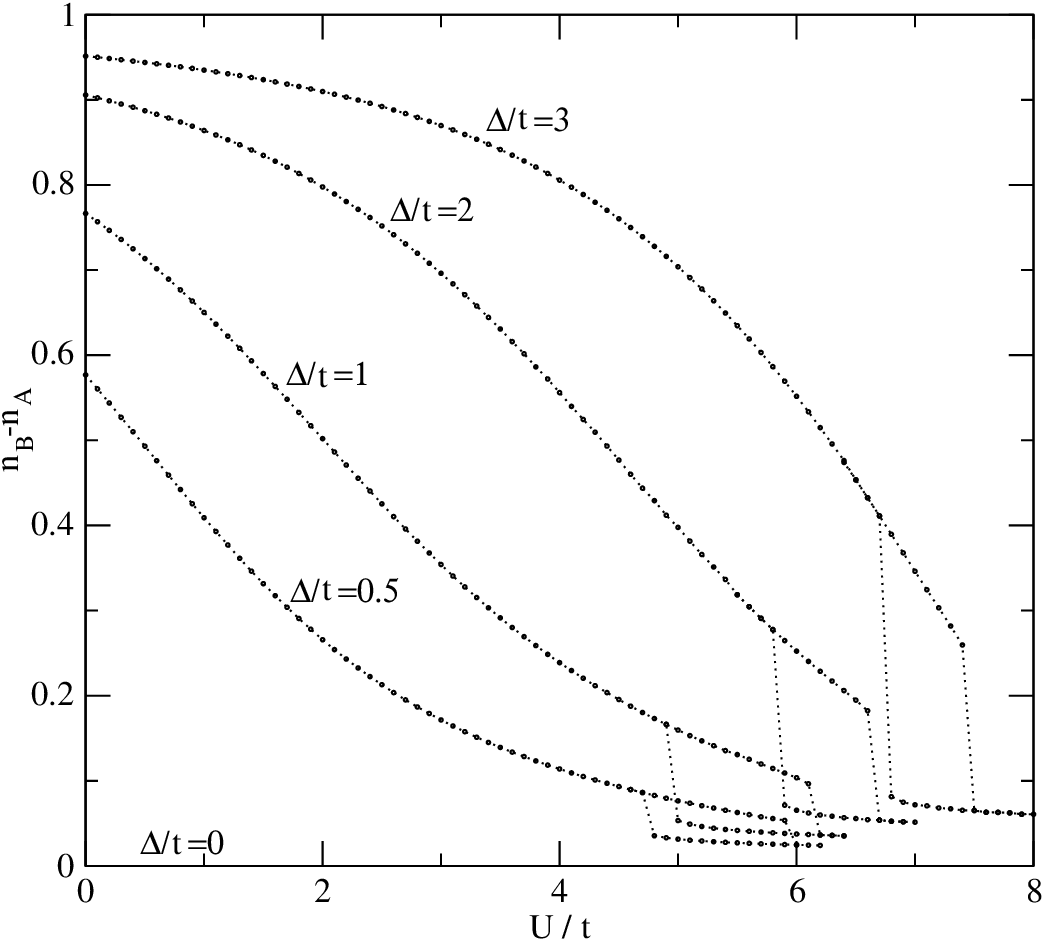}
\caption{Staggered charge density $n_B-n_A$ as a function of $U$ for
different values of the ionic potential $\Delta$. Our results for
$\Delta=2t$ are in good agreement with those obtained within cluster-DMFT
in 2D.~\cite{Kancharla}}
\label{stag}
\end{center}
\end{figure}

On a bipartite $A-B$ lattice with nearest-neighbor hopping $t$ the
single-site Green's functions reads
\begin{equation}
G_{\alpha}(\omega)=\xi_{\bar \alpha} \int_{-\infty}^{\infty} d
\varepsilon \frac{\rho_0(\varepsilon)}{\xi_A \xi_B -
\varepsilon^2} \;,
\label{GF}
\end{equation}
with $\alpha=A$ ($\bar \alpha=B$) and
$\xi_{\alpha}=\omega+i0^+-\varepsilon_{\alpha}+\mu-\Sigma_{\alpha}(\omega)$.
After integrating Eq.~(\ref{GF}) using the semi-circular DOS
$\rho_0(\varepsilon)=\sqrt{4t^2-\varepsilon^2}/(2\pi t^2)$, we
obtain two coupled equations
$G_{\alpha}(\omega)^{-1}=\xi_{\alpha}-t^2G_{\bar \alpha} (\omega)$
for the IHM on a Bethe lattice. As in Ref.~\onlinecite{Garg}, we use
the iterated perturbation theory (IPT)~\cite{KK96} to calculate the
local self-energies $\Sigma_{\alpha}(\omega)$. From a direct
comparison between different methods~\cite{KK96,Georges,xRG,QMC} it
is found for the one-band Hubbard model that the quasi-particle peak
(Kondo resonance), the position of the Hubbard bands, the
Mott-Hubbard gap, as well as the principal characteristics of the
metal-insulator transition in infinite dimensions are all well
captured by the IPT solver. Since at each lattice site the band
filling is always away from one-half in the IHM (see
Fig.~\ref{stag}), it is proper to use the canonical IPT {\it
ansatz}: $\Sigma_{\alpha}(\omega)=\Sigma_\alpha^{HF}+
A_\alpha\Sigma_\alpha^{(2)}(\omega)/(1-B_\alpha\Sigma_\alpha^{(2)}(\omega))$
for the correlated problem. Within this {\it ansatz}
$\Sigma_\alpha^{HF}=Un_\alpha$ is the first-order Hartree term and
$n_\alpha=-1/\pi\int_{-\infty}^0{\mathrm{Im}}G_\alpha(\omega)d\omega$.
The coefficients $A_\alpha$ and $B_\alpha$ are determined by
requiring $\Sigma_{\alpha}(\omega)$ to be exact at high-frequencies
and in the atomic limit. Finally, $\Sigma_\alpha^{(2)}(\omega)
\equiv \Sigma_\alpha^{(2)}(\omega,[\tilde {\mathcal
G}_{0\alpha}(\omega)])$ is the well known second-order diagram for
the self-energy.~\cite{KK96}
%

In our treatment, the host Green's functions $\tilde {\mathcal
G}_{0\alpha}$ required to compute $\Sigma_\alpha^{(2)} (\omega)$
self-consistently are written as $\tilde {\mathcal
G}_{0\alpha}^{-1}={\mathcal G}_{0\alpha}^{-1}
+\varepsilon_\alpha-\mu$. The energy shift $\varepsilon_\alpha-\mu$
introduced here accounts for the correct atomic limit of the IHM at
$U\gg \Delta,t$ (see below) as well as the mirror-type particle-hole
symmetry between sites $A$ and $B$. Notice that a similar type of
correction is employed in the context of single-site Hubbard model
($\varepsilon_A=\varepsilon_B=0$), where the energy shift is $-\mu$
($=-U/2$ at half filling).~\cite{Georges} In addition, we
point out that our energy shift is different from the Hartree
correction $-\Sigma_\alpha^{HF}$ employed in Ref.~\onlinecite{Garg},
and this might be the cause for significant differences between our
and their results 
in the large $U$ limit.

\begin{figure}[t]
\begin{center}\leavevmode
\includegraphics[width=8.0cm]{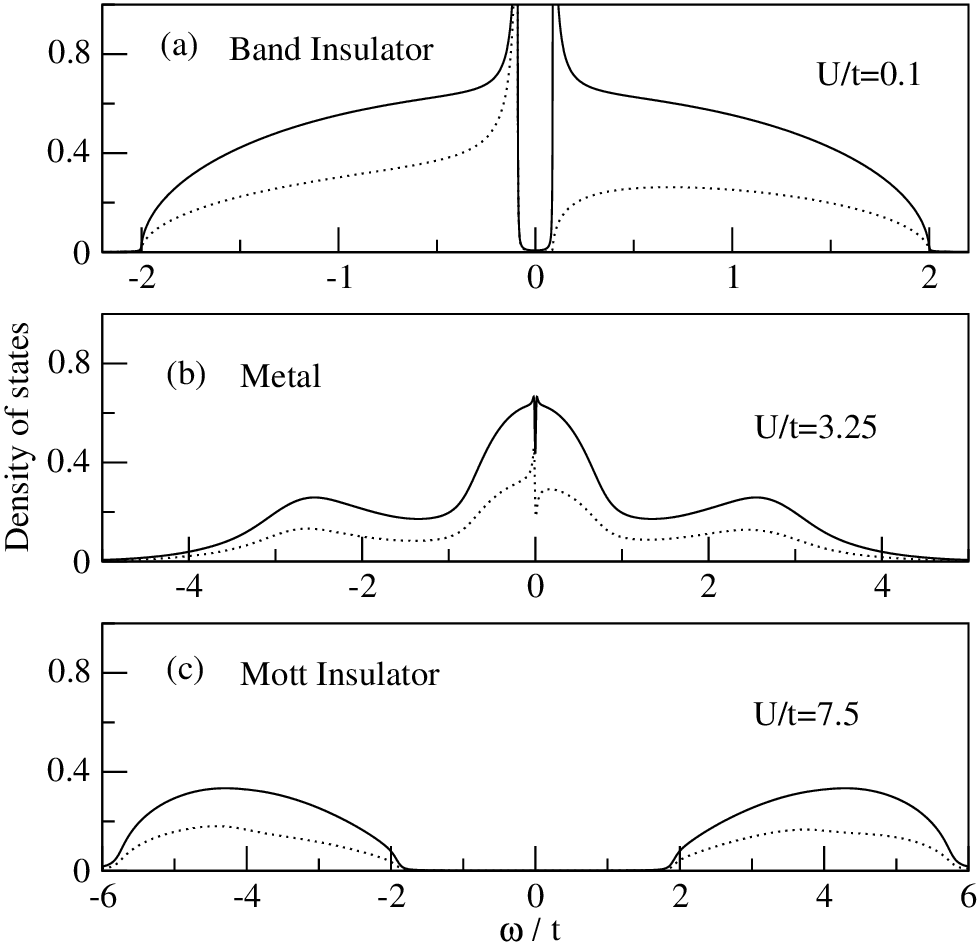}
\caption{Total and site selective (B) density of states (DOS) for
$\Delta=0.1t$ and various values $U$. The parameters are chosen
identical to those of Garg {\em et al.},~\cite{Garg} leading to very
close results besides for the lowest panel (larger gap and {\it
V-shape} leading edge in the DOS). Notice the small difference
between the lower and the upper Hubbard band in the site-selective DOS,
this results from the softening of the CDW at large $U$.}
\label{dos}
\end{center}
\end{figure}

\begin{figure}[t]
\begin{center}\leavevmode
\includegraphics[width=8.0cm]{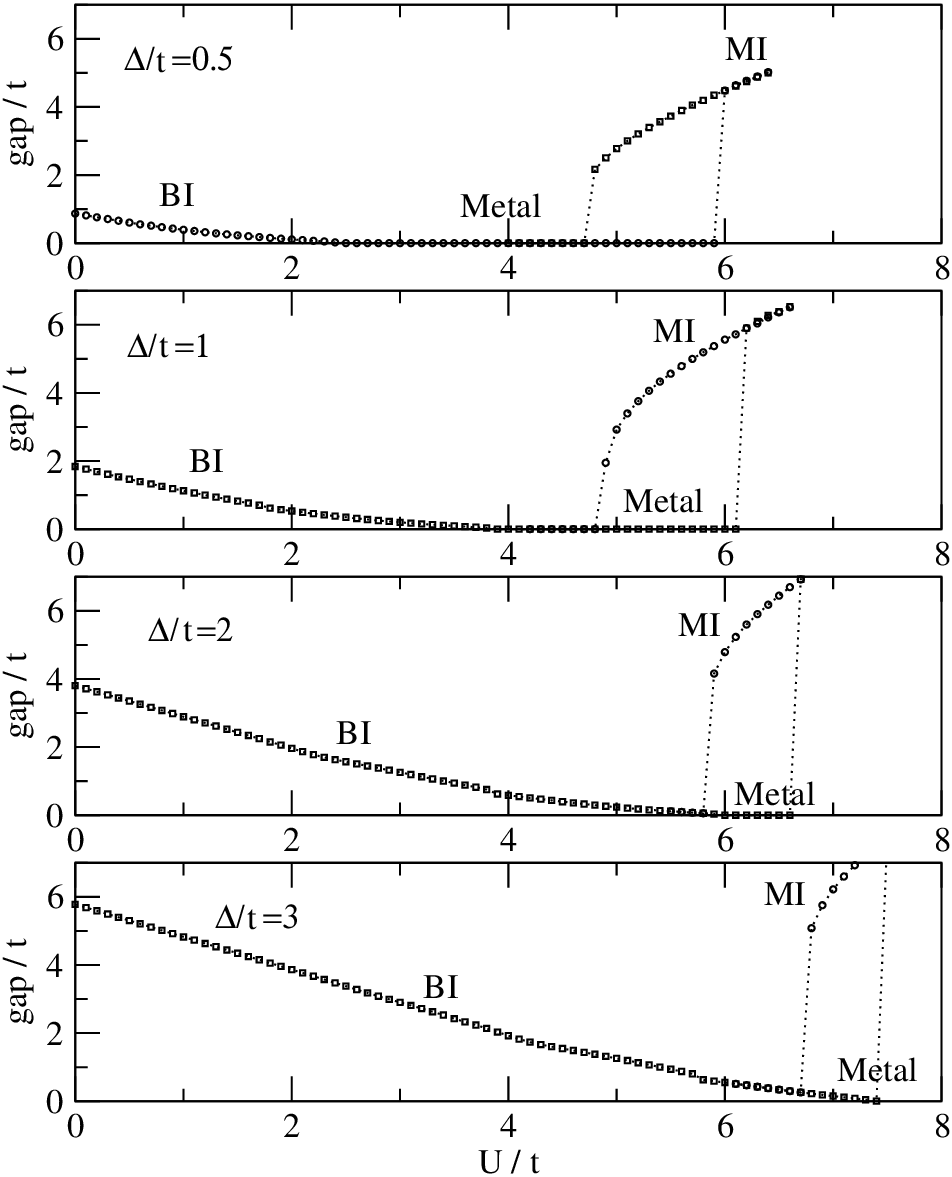}
\caption{Charge gap of the IMH as function of $U$
for various values of $\Delta$. The coexistence region between metal
and Mott insulator (MI) is reminiscent to DMFT for the pure Hubbard
model.~\cite{Georges} At large values of $\Delta$ the metallic phase
shrinks and, we obtain a coexistence region for the band (BI) and
Mott insulators.
}
\label{coex}
\end{center}
\end{figure}

Let us now present our results. In Fig.~\ref{dos}, we show the
spectral density for the charge phases of the IHM. In the upper
panel, we present our result for the BI phase for $U/t=0.1$. The
full line corresponds to the two-site averaged (or the {\it total})
DOS, whereas the dotted line is the site resolved B-DOS. As expected
the charge gap is small, of the order of $\Delta$. We observe clear
van-Hove singularities at the lower band edge at each $\alpha$-site.
With increasing $U$ the band gap is suppressed (by dynamical
transfer of spectral weight characteristic of correlated electron
systems) and a metallic phase sets in. As can be seen in the dotted
line of Fig.~\ref{dos}-(b), the B-DOS shows a pseudo gap feature at
the Fermi level. This implies that the one-particle DOS in the
metallic phase is not pinned to its unperturbed ($U=\Delta=0$)
value~\cite{MH}. This is understood, because the charge density wave
(CDW) regime prevents the Fermi liquid fixed point at any value of
the Coulomb interaction. As expected in such cases, the imaginary
part of $\Sigma_\alpha (\omega)$ (not shown) is finite at the Fermi
level. The dynamical transfer of spectral weight found in the
metallic phase can be traced to a Kondo like process that leads to
strong electron mass enhancement. A similar effect is known to take
place in correlated electron systems like the periodic Anderson
model~\cite{Georges} or in some regimes of the disordered Hubbard
model.~\cite{Tana} By further increasing $U$, the Mott-Hubbard
insulator with a gap of the order of $U-W$ ($W$ is the bandwidth)
sets in as the third phase of the IHM.

Having chosen exactly the same parameters as in Ref.~\onlinecite{Garg}
(and basically the same method of solution), we find very similar results
for the upper two panels, but substantial differences for the results in
the large $U$ regime are observed. In particular, the van-Hove structures
found in Ref.~\onlinecite{Garg} near the Fermi energy broadens and shifts
to high energies for $U=7.5t$. There is an appreciable modification of the
whole lineshape. Within our self-consistent scheme the Hubbard bands
centered at $\omega \approx \pm 4t$ are in very good agreement with those
found for the one-band Hubbard model. Clearly, our results for $U=7.5t$ are
more plausible than those obtained in Ref.~\onlinecite{Garg}.
Moreover, it is understood that in the large $U$ limit the staggered charge
density is small (see results of Fig.~\ref{stag}), forcing one to recover
to a good extent the physics of the regular Hubbard model. This is exactly
what is shown in the lower panel of Fig.~\ref{dos}.

\begin{figure}[t]
\begin{center}\leavevmode
\includegraphics[width=8.0cm]{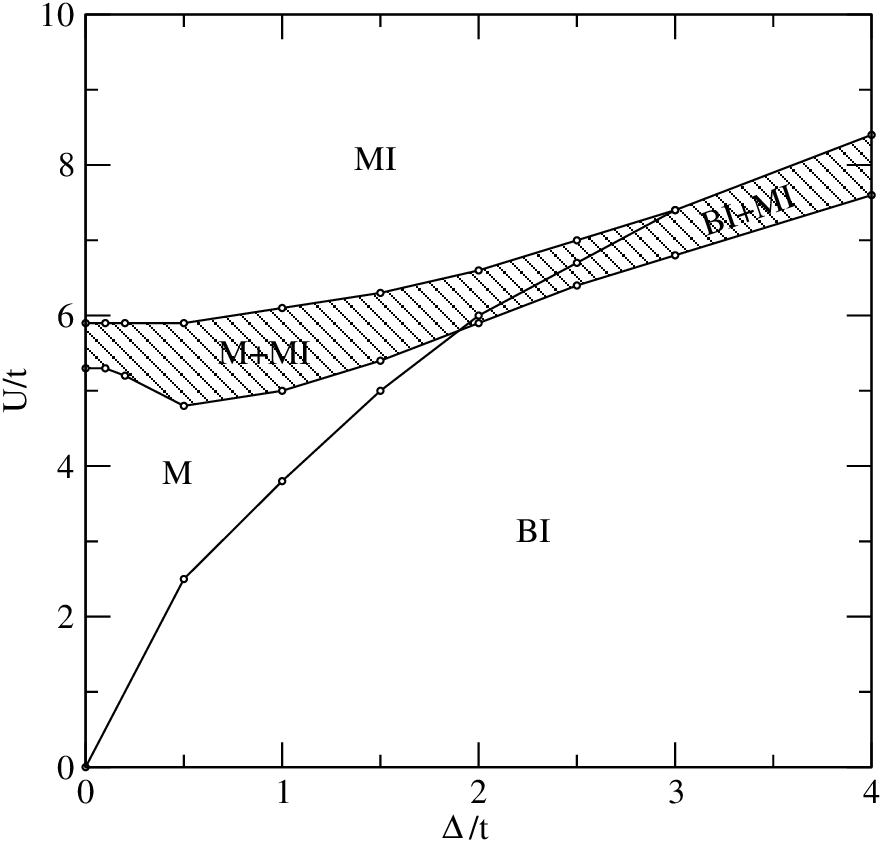}
\caption{Phase diagram of the IHM 
obtained by single-site DMFT. The dashed area corresponds to the
coexistence region between metal (M) and Mott insulator (MI) or
between band insulator (BI) and Mott insulator, respectively. The
line separating the BI and M phase is characterized by the vanishing
of the charge gap.} \label{pd}
\end{center}
\end{figure}

To characterize the correlated nature of the various phase
transitions and the stability of the metallic phase, we show in
Fig.~\ref{coex} the obtained charge gap as a function of $U$. As can
be seen in this figure, the width of the metallic phase shrinks
until it disappears at a critical value. For $\Delta\leq
1.9 t$, we find a continuous transition from the BI to a
unique metallic phase with increasing $U$. Differently from the
BI-to-metal transition, the transition from the metallic phase to
the Mott insulator phase is of the first order. The latter
transition in the IHM is completely identical to the corresponding
transition in the ordinary Hubbard model. Across the metal-to-MI
transition line we observe a coexistence region consisting of two
distinct (numerical) solutions, which is accompanied by a hysteresis
loop characteristic of a first order MIT.~\cite{Georges} In
agreement with previous theoretical studies,~\cite{Garg,Kancharla}
the width of the coexistence region decreases with increasing
ionicity ($\Delta$). For larger values of $\Delta$, the metallic
phase disappears and a coexistence region between BI and MI is
found.

The obtained phase diagram for the IHM on a Bethe lattice is shown
in Fig.~\ref{pd}. It differs considerably from those found and/or
recently proposed for the IHM on a usual bipartite lattice in two or
at high dimensions.~\cite{Garg,Paris,Kancharla,Bouadim} Differently
from previous single-site DMFT results.~\cite{Garg} the metallic
phase is now considerably widened up to $1.9t$. The
first-order metal-to-MI and BI-to-MI phase transitions resemble the
coexistence region found by Kancharla {\it et al.} in
2D.~\cite{Kancharla} In addition, our window for the metallic region
at moderate values of $\Delta$ is similar to the determinant quantum
Monte Carlo study also in 2D,~\cite{Paris,Bouadim} indicating common
similarities between the physics at low and high dimensions for the
IHM. To further illustrate the underlying features in the
coexistence region and the first-order phase transitions, we show in
Fig.~\ref{stag} the staggered  charge density $n_B-n_A$ as a
function of $U$. As one can see, there is no peculiarity at the
transition point between the BI and the metallic phase. However, the
staggered charge density jumps at the transition towards the MI
phase.

The possibility of having different types of first-order phase
transitions in the IHM (see Fig.~\ref{pd}) is one relevant aspect of
our work. Furthermore, our results show that the coexistence region
is not necessarily connected with the bond-order phase recently
found with cluster-DMFT for the 2D-IHM.~\cite{Kancharla} We find
that a first-order transition between metal and MI (or between BI
and MI) exists not only in 2D but also at high dimensions. Thus, the
first-order MIT of the IHM is of the same nature as that found
within the single-site DMFT for the one-band Hubbard
model.~\cite{Georges}
Finally, we recall that the 2D~\cite{Bouadim,Kancharla} works reach
opposite conclusions regarding the metallic phase of the IHM,
indicating a great sensitivity on the procedure used. The metallic
phase derived within determinant quantum Monte Carlo~\cite{Paris}
suggests a {\it quasi-particle peak} at $\omega=0$,~\cite{Bouadim}
while a {\it pseudo-gap} like feature is seen in the Lanczos
solution of the cluster bath.~\cite{Kancharla} Clearly, near the
Fermi level our metallic phase is more consistent with the CDMFT
calculations. However, the shape of our phase diagram is similar to
the one derived in Ref.~\onlinecite{Paris} which indicates that the
correct nature of the metallic phase in 2D remains to be
investigated.

From the pure Hubbard model it is known that IPT method is well
suited to show the existence of the coexistence region and provides
a correct qualitative picture of the metal insulator transition in
infinite dimensions. However, there are some details of this transition, like e.g.
the precise position or the stability of different phases~\cite{Georges},
which have to be clarified by more sophisticated methods
and which are left for further studies.

In summary, we derived the many-body phase diagram of the IHM in the
limit of infinite dimensions. For intermediate values of the ionic
potential $\Delta$, we find a metallic phase between the BI and the
Mott-Hubbard insulating phase which disappears, however, for larger
$\Delta$. Our results clearly differ from a previous work which
employs the same single-site approach for the IHM.~\cite{Garg}
We believe to have found the correct spectral
density in the limit of $U\gg \Delta$. Differently from
Ref.~\onlinecite{Garg}, our result in this regime shows clear
Mott-Hubbard features, in agreement with the general
expectation.~\cite{Georges} We find first-order phase transitions
between the metallic and the MI phases as well as between the band
and the Mott insulator. The latter one were not reported in previous
studies on metal-insulator transitions for the IHM. We show that a
discontinuous transition exists already at high dimensions and is
not necessarily connected with the bond-order phase as proposed in
Ref.~\onlinecite{Kancharla}. 


R.H. thanks the NATO science division (CLG 98 1255) for financial
support. L.C. acknowledges support from the SFB608 and the Emmy
Noether-Programm of the DFG. He also thanks M.S. Laad for discussions
as well as the University Paul Cezanne at Marseille for hospitality.
G.I.J. acknowledges support from the STCU-grant N 3867 and from the
Georgian National Science Foundation grant GNSF/ST06/4-018.

\end{document}